\documentclass[twocolumn,showpacs,preprintnumbers,amsmath,amssymb]{revtex4}

\usepackage{graphicx}% Include figure files
\usepackage{dcolumn}% Align table columns on decimal point
\usepackage{bm}% bold math

\begin{document}

\title{Central depression in nuclear density and its consequences 
for the shell structure of superheavy nuclei.}

\author{A.\ V.\ Afanasjev$^{(1,2)}$, S.\ Frauendorf$^{(1,3)}$}

%\author{A.\ V.\ Afanasjev}
%\affiliation{Department of Physics, University of Notre Dame,
%Notre Dame, Indiana 46556, USA}
%\affiliation{Laboratory of Radiation Physics, Institute of Solid State
%Physics, University of Latvia, LV 2169 Salaspils, Miera str. 31, Latvia}
%\author{S.\ Frauendorf}
%\affiliation{Department of Physics, University of Notre Dame,
%Notre Dame, Indiana 46556, USA}
%\affiliation{IKH, Research Center Rossendorf, Dresden, Germany}

\address{$^{1}$Department of Physics, University of Notre Dame,
Notre Dame, Indiana 46556, USA}
\address{$^{2}$Laboratory of Radiation Physics, Institute of Solid State
Physics, University of Latvia, LV 2169 Salaspils, Miera str. 31, Latvia}
\address{$^{3}$IKH, Research Center Rossendorf, Dresden, Germany}

\date{\today}

\begin{abstract}

     The influence of the central depression in the density distribution of
spherical superheavy nuclei on the shell structure is studied within the 
relativistic mean field theory. Large depression leads to the shell gaps at 
the proton $Z=120$ and neutron $N=172$ numbers, while flatter density distribution 
favors $N=184$ for neutrons and leads to the appearance of a $Z=126$ shell gap 
and to the decrease of the size of the $Z=120$ shell gap. The correlations 
between the magic shell gaps and the magnitude of central depression are 
discussed for relativistic and non-relativistic mean field theories.
\end{abstract}
                                                                                
\maketitle

  The question of the possible existence of shell-stabilized superheavy 
nuclei and the precise location of the magic spherical superheavy nuclei 
has been in the focus of the nuclear physics community for more than three 
decades \cite{HM.00}. Unfortunately, the various  theoretical models do not agree 
with respect to the magic shell gaps in superheavy nuclei. 
The proton numbers  $Z=114, 120$ and 126 and the neutron numbers $N=172$ 
and 184  are predicted by different models and parametrizations 
\cite{RBM.02}.
The same models reproduce the known magic numbers 
in lighter systems. The predicted magic numbers are of decisive importance for
the   experimental search of superheavy nuclei.
In such a situation it is necessary to understand what 
makes the predictions of the  models so different in the region 
of superheavy nuclei. One of the reasons is the appearance of a 
central depression in the nuclear density \cite{BRRMG.99,DBDW.99}, 
which is studied in this paper.

 The first predictions of superheavy nuclei were based on the 
shell correction method, which assumes a single-particle potential
with a flat bottom. Nowadays, these 
calculations  predict $Z=114$ and N=184. The microscopic self-consistent 
models start either from effective nucleon-nucleon interaction (models 
based on the Skyrme and Gogny forces) or effective exchange of mesons 
by nucleons (relativistic mean field (RMF) theory). 
 Although based on more fundamental principles,
these models do not agree among each other in predicting
the  magic shell gaps of superheavy nuclei. In part, this is related
to the fact that the reliability of different parameterizations of 
these models is verified only by comparing theoretical and experimental
binding energies and its derivatives (separation energies, the 
$\delta_{2n,2p}(Z,N)$ quantities) and deformation properties \cite{AKFLA.03}. 
These observables are not very sensitive to the energies of the 
single-particle states. For example, it was shown in Ref.\ \cite{AKFLA.03} 
that the NLSH and NL-RA1 parameterizations of the RMF theory provide a 
reasonable description of these quantities in deformed actinide region 
despite the fact that the single-particle energies are poorly reproduced. 
Accurate single-particle energies 
are crucial for predicting the shell gaps in superheavy nuclei. However, 
the accuracy of the description of the single-particle states 
in deformed region of the heaviest actinides has been tested only for 
few parameterizations of the RMF theory \cite{AKFLA.03} and Skyrme SLy4 
functional \cite{BBDH.03}. The shell correction approach is based on
phenomenological potentials which reproduce the single particle levels
of the actinides best. However, the assumption of a flat bottom 
radial profile is a severe source of error when extrapolating
to spherical superheavy nuclei.

  Self-consistent microscopic calculations find a central depression in the 
nuclear density distribution \cite{BRRMG.99,DBDW.99}, which generates 
a wine-bottle nucleonic potential. Its magic numbers differ 
from the ones of the  phenomenological  flat-bottom potentials. The present 
manuscript studies the influence of this depression on 
the shell structure of spherical superheavy nuclei. As a theoretical tool 
we use the RMF theory for spherical nuclei without pairing \cite{R.96} and 
the relativistic Hartree-Bogoliubov (RHB) theory \cite{ARK.00}.

%------------------------------------------------------------
\begin{figure}
\includegraphics[angle=0,width=8.5cm]{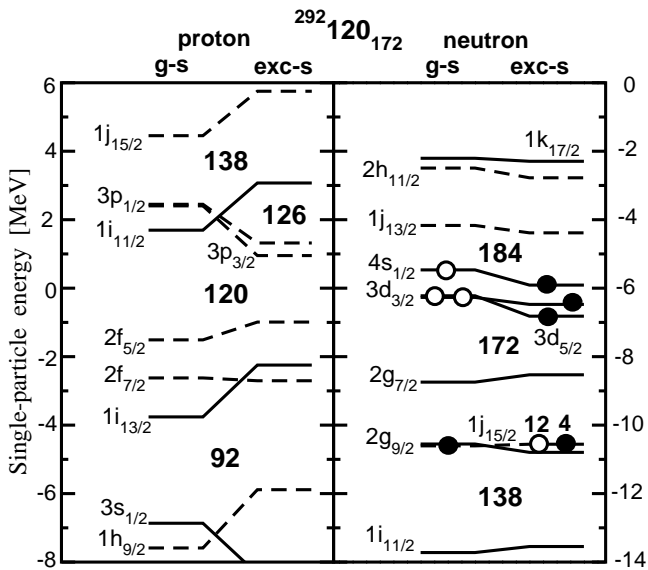}
\caption{\label{z120n172-sp}Single-particle spectra of ground state 
(indicated as 'g-s') and excited (indicated as 'exc-s') configurations 
in $^{292}120_{172}$ system obtained in the RMF calculations with the 
NL3 force. Solid and dashed lines are used for positive and negative 
parity, respectively. Solid and open circles indicate the occupied and 
empty orbitals, respectively. In the ground state, all subshells below 
$Z=120$ and $N=184$ are fully occupied. For the g-s configuration, 
the spin-orbit partners $3p_{1/2}, 3p_{3/2}$ and $3d_{3/2}, 3d_{5/2}$ 
show up at very close energy. In the excited configuration, only 12 
particles are excited from 
the subshell $\nu 1j_{15/2}$: 4 particles still reside in this 
subshell. The spherical shell gaps of interest are indicated.}
\end{figure}
%-------------------------------------------------------------

  Fig.\ \ref{z120n172-sp} compares the single-particle spectrum of a
wine-bottle potential (g-s) with the one of a flat-bottom 
potential (exc-s). The details of the potentials will be discussed later.  
The differences are easy to understand. The high-$j$ orbitals are 
localized mostly near the surface, whereas the low-$j$ orbitals have a 
more central localization. As compared to a flat-bottom potential,
in general the high-$j$ orbitals are more and the low-$j$ orbitals are 
less bound in an attractive wine-bottle potential. In the following 
it is useful to distinguish between the groups of low-$j$ 
and high-$j$ single-particle states. Filling up a low-$j$ group with 
nucleons increases the density near the center, whereas filling a high-$j$ 
group increases the density near the surface. As we shall demonstrate, 
the occupation of these groups determines the radial profile of the 
neutron and proton densities and potentials.

%------------------------------------------------------------
\begin{figure*}
\includegraphics[angle=0,width=13.5cm]{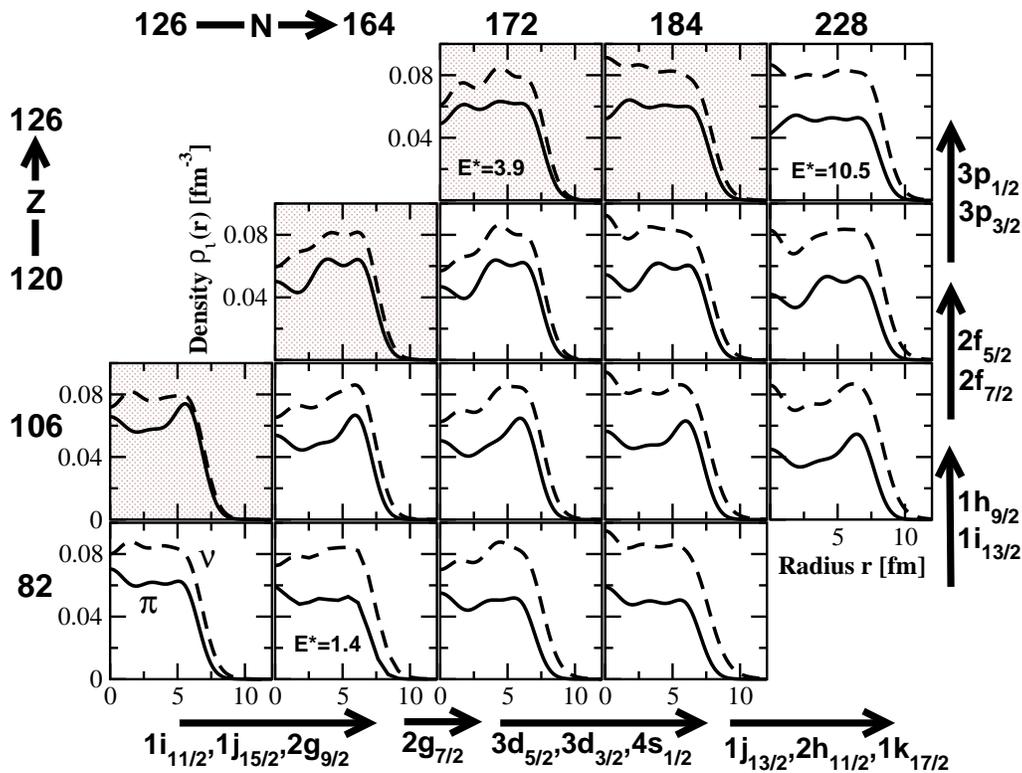}
\caption{\label{dens-sys} The evolution of proton and neutron densities 
with the changes of proton and neutron numbers. Arrows indicate the group 
of single-particle subshells which become occupied with the change 
of the nucleon number. The figure is based on the results of spherical 
RMF calculations without pairing employing the NL3 parametrization. The 
shaded background is used for nuclei located beyond the 
proton-drip line. If the indicated configuration is not lowest in energy, 
its excitation energy (in MeV) is given by E*.}
\end{figure*}
%-------------------------------------------------------------

  We start with $^{208}$Pb. The RMF theory provides a good description of 
the experimental charge density distribution of 
this nucleus \cite{GRT.90,NL1}. With 
increasing neutron and proton numbers the corresponding densities
are modified in the way shown in Fig.\ \ref{dens-sys}. 
Between  $Z=82$ and $Z=106$ the protons fill the high-$j$ group 
 $\pi 1i_{13/2}$, $\pi 2h_{9/2}$,  between $Z=106$ and $Z=120$ they fill the
 medium-$j$ group $\pi 2f_{7/2}, \pi 2f_{5/2}$, and between $Z=120$ and $Z=126$ 
they fill the low-$j$ group $\pi 3p_{1/2}, \pi 3p_{3/2}$. The  variations of 
the proton density  are seen in the $N=172$ isotones. The filling of the 
high-$j$ group $\pi 1i_{13/2}, \pi 2h_{9/2}$ increases the density at the 
surface (compare $Z=82$ and $Z=106$ in Fig.\ \ref{dens-sys}). The 
filling of medium-$j$ group $\pi 2f_{7/2}, \pi 2f_{5/2}$ adds to 
the density between central and surface areas (see 
$Z=120$). Finally, the filling of the low-$j$ group $\pi 3p_{1/2}, 
\pi 3p_{3/2}$ adds to the density in the near-central 
region of nucleus (see $Z=126$).

  The analogous grouping into low-$j$ and high-$j$ subshells in the neutron 
subsystem is illustrated in Fig.\ \ref{dens-sys}. The variation of the neutron 
density generated by filling these groups is seen most clearly in the $Z=106$ 
isotopes. Filling the high-$j$ group $\nu 1i_{11/2}, \nu 1j_{15/2}, \nu 2g_{9/2}$ 
increases the density near the surface. Filling the low-$j$ group 
$\nu 3d_{5/2}, \nu 3d_{3/2}, \nu 4s_{1/2}$ increases the central density 
and filling the high-$j$ group $\nu 1j_{13/2}, \nu 2h_{11/2}, \nu 1k_{17/2}$ 
adds to the surface region again. Analyzing the published results, we found that 
the grouping into high/medium/low-$j$ subshells shown in Fig.\ \ref{dens-sys} 
appears in all models/parametrizations (cf. Fig.\ \ref{z120n172-sp} in the 
present manuscript and Figs.\ 4, 9, 13, and 15 in Ref.\ \cite{BRRMG.99}).

  As seen in  Fig.\ \ref{dens-sys},  the combined  occupation of the high-$j$ 
neutron subshells $2g_{9/2}$, $1j_{15/2}$, $1i_{11/2}$ [and medium-$j$ $2g_{7/2}$] 
and proton $1h_{9/2}$ and $1i_{13/2}$ [and medium-$j$ $2f_{7/2}$] subshells 
leads to a central depression in the nuclear density 
between $Z=106$ and $Z=120$ and $N=164$ and $N=172$, which is especially 
pronounced in the $Z=120,N=172$ system. As seen from the density
variations in Fig.\ \ref{dens-sys}, 
the proton subsystem plays a larger role in the creation of the central depression. 
This result differs from the results of the Skyrme calculations with the SkI3 
parametrization \cite{BRRMG.99}, the authors of which claim that the 
central depression is  mainly due to the occupation 
of neutron subshells. The appearance of the central depression is a consequence 
of the different density distributions of the single-particle states: high-$j$ 
orbits are located near the surface and low-$j$ orbits near the center.  
This generic feature is dictated by the nodal structure of the wave 
functions in a leptodermic potential. Hence, the high-$j$ proton and neutron 
orbitals will modify the radial profile in a comparable way. However, the high-$j$ 
proton orbitals should be more efficient, because  the Coulomb interaction pushes 
them to larger radii.

%------------------------------------------------------------
\begin{figure}
\includegraphics[angle=0,width=8cm]{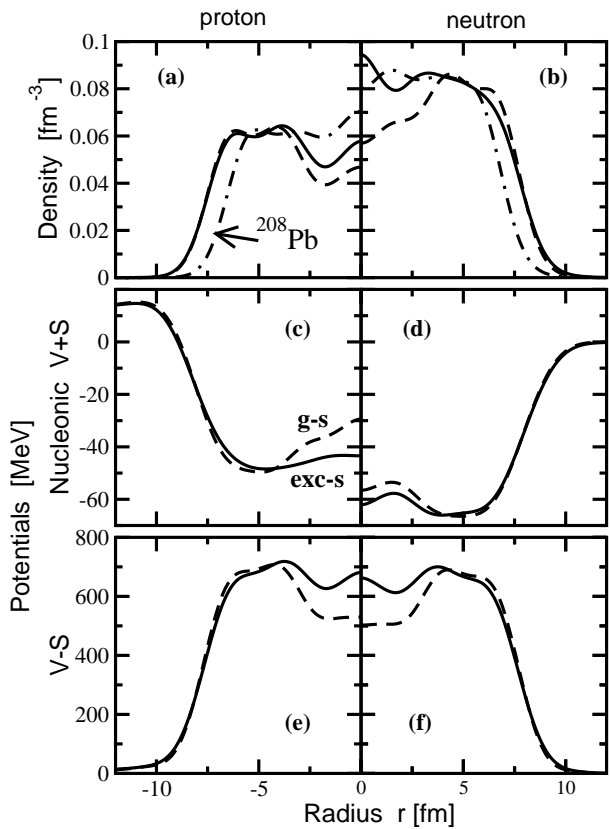}  
\caption{\label{dens-pot-z120n172} Density distributions (upper row), 
nucleonic ($V+S$, middle row) and  $V-S$ (bottom row) potentials 
in the ground (g-s) and excited (exc-s) configurations of $^{292}120_{172}$.
$S$ and $V$ are attractive scalar and repulsive vector potentials,
respectively. The left column shows 
the proton system and the right the neutron system. The proton and neutron 
density distributions of $^{208}$Pb are shown in upper panels for comparison.}
\end{figure}
%-------------------------------------------------------------

%------------------------------------------------------------
\begin{figure}
\includegraphics[angle=0,width=7cm]{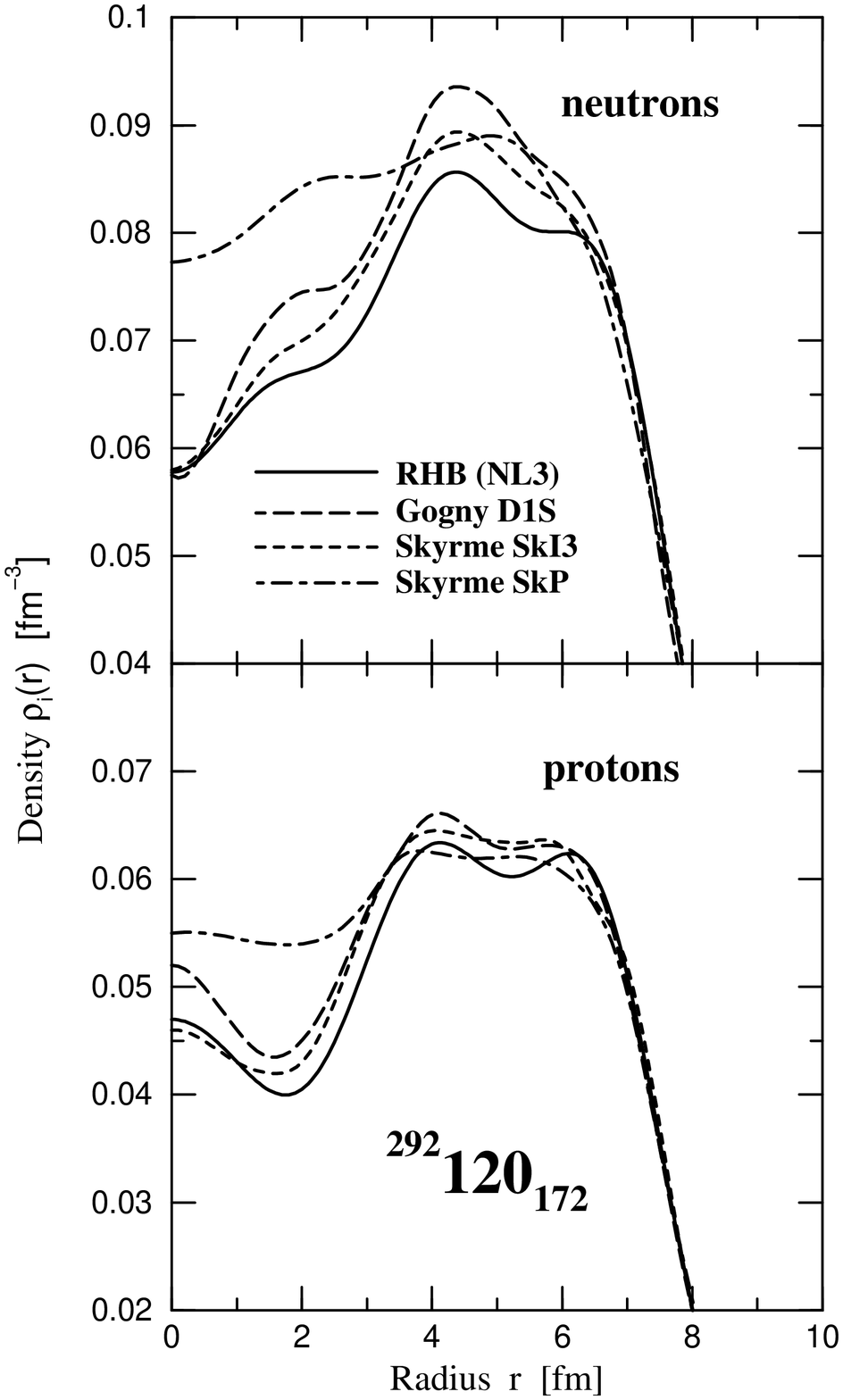}  
\caption{\label{dens-oth-mod}
    Neutron and proton densities of $^{292}120_{172}$ obtained 
in different models/parametrizations. The densities obtained with Skyrme 
and Gogny forces are taken from Refs.\ \protect\cite{BRRMG.99,BBDGD.01}.}
\end{figure}
%-------------------------------------------------------------

Let us study the interplay between the geometry of the single-particle orbitals, 
the appearance of the central depression in the density and the shell structure 
in more detail. One possibility is to generate a flatter density distribution 
in the central part of nucleus by exciting particles 
from high-$j$ subshells to low-$j$ subshells. 
Fig.\ \ref{z120n172-sp} shows an example of such an excitation in the 
$^{292}120_{172}$ system. Here 12 neutrons are excited from the 
$1j_{15/2}$ subshell into $3d_{5/2}$, $3d_{3/2}$ and $4s_{1/2}$ subshells. 
In this excited (called 'exc-s') configuration the neutron density distribution 
in the central part of nucleus is much flatter than in the ground state (called 
'g-s') configuration, and its profile is very similar to the one in $^{208}$Pb 
(Fig.\ \ref{dens-pot-z120n172}b). The changes in neutron density are fed back 
to proton density, because the isovector interaction tries
to keep them alike. As a consequence, the proton density distribution becomes 
also flatter (Fig.\ \ref{dens-pot-z120n172}a) but  density fluctuations due to 
shell effects remain visible.  The nucleonic $V+S$ (Fig.\ \ref{dens-pot-z120n172}c,d)
potential and the $(V-S)$ (Fig.\ \ref{dens-pot-z120n172}e,f) potential,
which in first approximation is related to the spin-orbit potential 
via $V_{ls}({\bf r})=\frac{m}{m^*({\bf r})}(V({\bf r})-S({\bf r}))$ \cite{Vls},
reflect the density change: they becomes flatter in the central part of nucleus. 
This effect is especially pronounced in the proton nucleonic potential: the 
'wine-bottle' radial shape is replaced by a 'flat-bottom' one. Another consequence 
of this excitation is an increase of the surface diffuseness both in the densities 
and in the potentials.

The various RMF forces are characterized by different compression moduli $K_{\infty}$ 
(NL-Z [$K_{\infty}=173$ MeV], NL3 [$K_{\infty}=272$ MeV] and NLSH 
[$K_{\infty}=355$ MeV]). As expected, the magnitude of central depression in densities 
and potentials increases with the decrease of compression modulus. However, 
the changes in the energies of single-particle states, densities and potentials induced 
by our  probing particle-hole excitation do not depend sensitively on the compressibility.

  As a result of the flattening  of the nucleonic potential the energies of the 
single-particle states are changed as described above (see Fig.\ \ref{z120n172-sp}). 
The shifts are larger in the proton subsystem because the proton central potential 
is more flattened than the neutron one.  We see the  $Z=126$ proton gap emerging 
and the size of the $Z=120$ gap decreasing. To a lesser extend, the   $N=172$ 
neutron gap decreases and the $N=184$ gap increases. The flattening of the $(V-S)$ 
potential increases the splitting of the  spin-orbit pairs  [$\pi 3p_{1/2}$, 
$\pi 3p_{3/2}$], 
[$\nu 3d_{3/2}$, $\nu 3d_{5/2}$], and 
[$\pi 2f_{5/2}$ and $\pi 2f_{7/2}$]. The spin-orbit splitting of the
last pair of orbitals generates the $Z=114$ shell gap predicted
by a number of models. The present results clearly show that a flatter 
density distribution leads to a larger splitting between these orbitals.
 
 We have studied further excitations that induce a flatter density 
distribution. In all cases we found the above-mentioned dependence 
of the size of the $Z=120,126$ and $N=172,184$ shell gaps on the magnitude 
of the central density depression. Our results are consistent with the HFB 
studies with the Gogny D1S force, which employ an external potential in order 
to induce the central depression \cite{DBDW.99}: large $N=184$ and $Z=126$ 
shell gaps were found for the values of the external potential that generate 
a flat density distribution  and large $Z=120$ 
and $N=172$ shell gaps for the values that generate  a central depression (see 
Fig.\ 2 in Ref.\ \cite{DBDW.99}).

 Due to the isovector force, which tries to keep the neutron and 
proton density profiles alike, there  is a mutual enhancement of the 
$Z=120$ and $N=172$ gaps, both being favored by the wine-bottle potential,
and of the $Z=126$ and $N=184$ gaps, both favoured by the flat bottom 
potential. For the same reason the gaps are smaller for the combination 
$Z=126$ and $N=172$, and the $Z=120$ gap does not develop for $N=184$.  
This behavior is not expected to depend much on the 
density functional chosen. Indeed, a number of Skyrme calculations 
(SkI3, SkI4, SkI1, SLy6), which show a large $Z=120$ gap in the $^{292}120_{172}$ 
system, do not show double shell closure at $Z=120,N=184$ \cite{RBBSRMG.97}.
These generic features are also seen in the calculations with Gogny D1S force
(Fig.\ 2 of Ref.\ \cite{BBDGD.01}),  with the SkI1 (Fig.\ 2 in Ref.\ \cite{RBBSRMG.97}),  
and SkI3, SkI4, SkP (Figs.\ 6, 7, and 8 in Ref.\ \cite{BRRMG.99}) Skyrme
forces, and with the RMF NL3 and NL-Z2 forces (Fig.\ 2 in Ref.\ \cite{KBNRVC.00}). 

    Let us consider the $^{292}120_{172}$ system, which is a doubly magic superheavy 
nucleus in RMF theory. Both relativistic and non-relativistic (Gogny D1S, Skyrme 
parametrizations with low isoscalar effective mass $m^*/m$ such as SkI3, 
SLy6 \cite{BRRMG.99}) show a pronounced central depression (see Fig.\ 
\ref{dens-oth-mod}). These density functionals are characterized by similar 
values of $m^*/m$ (Gogny D1S [$m^*/m=0.67$], Skyrme SkI3 [$m^*/m=0.57$] and SLy6 
[$m^*/m=0.69$] \cite{BRRMG.99}. These values should be compared with RMF Lorentz 
effective mass of the nucleon at the Fermi surface $m^*(k_F)/m\approx 0.66$ 
\cite{BRRMG.99}, since effective mass is momentum-dependent  in the RMF theory
\cite{JM.89}. The central depression is much smaller in the Skyrme calculations 
with SkP (Fig.\ \ref{dens-oth-mod}) and SkM*  forces (Ref.\ \cite{BRRMG.99}) which 
have high values of isoscalar effective mass $m^*/m=1$ and $m^*/m=0.789$, respectively. 
The development of a more pronounced central depression for the density functionals with
low effective mass may be understood as follows. In the surface region, the ratio
$m^*/m$ changes from its value $<1$ in the interior to 1 in the exterior. Classically, 
nucleons with given kinetic energy are more likely to be found in regions with high 
effective mass than
in the regions with low one because they travel with lower speed. This is reflected by the 
Thomas-Fermi expression for the nucleonic density 
$\rho \propto [2 m^*(\epsilon_F-V)]^{3/2}$.
The increase of the effective mass in the surface
region favours the transfer of mass from the center there, which makes the above discussed
polarization mechanism of the high-$j$ orbitals more effective for functionals
with low effective mass.
Based on this argument we suggest that a flatter radial profile is a generic feature of 
the density functionals with an effective mass close to one.
It would be interesting to investigate if the Skyrme functionals of this type 
systematically give flatter density distributions than the ones with a 
low effective mass.

   All experimentally known nuclei with $Z\geq 100$ are expected to be deformed 
\cite{HM.00,O.01}. The deformation leads to a more equal distribution of the 
single-particle states emerging from the high-$j$ and low-$j$ spherical subshells 
(see, for example, the Nilsson diagrams in Figs.\ 3-4 in Ref.\ \cite{CAFE.77})
than for spherical shape. Thus, the density profile of deformed nuclei is relatively 
flat \cite{AF.lim04}, strongly resembling the one used in phenomenological potentials. 
This together with a careful fit of the single-particle energies to the deformed 
nuclei in heavy actinide region explains the success of the shell correction method 
\cite{MN.94,PS.91}. However, this method neglects the self-consistent rearrangement 
of single-particle levels due to the appearance of a central depression  in spherical 
superheavy nuclei. Thus predictions of magic numbers for superheavy nuclei within 
the shell correction method should be  considered with caution.

   We have deliberately excluded from our study the forces NLSH and NL-RA1 [RMF] 
and SkI4 [Skyrme], which give a $Z=114$ shell gap in self-consistent calculations. 
This is because they provide a poor description of either the energies of single-particle 
states in deformed actinide $A\sim 250$ nuclei \cite{AKFLA.03,B.03} or of the 
spin-orbit splitting \cite{BRRMG.99}. The energy splitting between deformed states 
emerging from the $2f_{7/2}$ and $2f_{5/2}$
subshells is well described by the RMF NL1 and NL3 \cite{AKFLA.03} and Skyrme SLy4 
\cite{BBDH.03} forces, which give a small $Z=114$ shell gap. 
In our opinion, these results make the predicted shell gaps at $Z=120,126$ and 
$N=172,184$ most likely. As discussed above, their 
appearance and combination depends on the magnitude of the central depression.  
The RMF theory gives  a pronounced double shell closure at $Z=120,N=172$.  
The non-relativistic theories (Gogny \cite{BBDGD.01}, Skyrme \cite{RBBSRMG.97,KBNRVC.00}) 
give a large shell gap at $N=184$ and less pronounced gaps at $Z=120$ and 126, 
the size of which strongly depends on neutron number. For example,
the Skyrme forces with high effective mass (SkM*, SkP) tend to predict a double 
shell closure at $Z=126,N=184$, while those with low effective mass 
(SkI1, SkI3, SkI4, SLy6) show a large gap at $Z=120$ for $N=172$, which becomes 
smaller or disappears when approaching $N=184$.

   In RMF theory, the $N=172$ gap lies between the subshells $\nu 3d_{5/2}$ and 
$\nu 2g_{7/2}$, which form a pseudospin doublet \cite{KBNRVC.00}. The 
analysis of their deformed counterparts in the $A\sim 250$ region shows that the 
experimental energy distance between the pseudospin partners $\nu 1/2[620]$ and 
$\nu 3/2[622]$ is well reproduced, which supports the predicted existence of a 
gap at $N=172$ (see Fig.\ 28 in Ref.\ \cite{AKFLA.03}). However, taking into account 
the typical uncertainty of the description of the single-particle states in 
best-tested RMF parametrizations, one cannot exclude a large gap at 
$N=184$ \cite{AKFLA.03}. For this to take place, the energy of $\nu 4s_{1/2}$ 
state has to be overestimated by approximately 1 MeV.

    In summary, the influence of the filling of the spherical subshells on
the radial density profile and shell structure of superheavy nuclei has 
been studied. The occupation of high-$j$ subshells decreases the density 
in the central part of the nucleus, the occupation of low-$j$ subshells 
increases it. The polarization due to high-$j$ orbitals generates a 
central depression of the density for nuclei with $Z\approx 120$ and/or 
$N\approx 172$, which is particularly pronounced for the combination 
$Z=120,N=172$, because both the proton and the neutron subsystems induce 
a central depression. This large central depression produces large shell gaps
at $Z=120$ and $N=172$. The occupation of low-$j$ orbitals by means of either 
multi-particle-hole excitations or of the increase of $Z,N$ beyond $Z=120,N=172$ 
removes the central depression and reduces these 
shell gaps. The shell gaps at $Z=126$ and $N=184$ are favored 
by a flat density distribution in the central part of nucleus. The magnitude 
of central density  depression correlates also with the effective mass of 
nucleons: low effective mass favors the large central depression. 
The similarties and differences between non-relativistic and relativistic 
mean field models in the predictions of magic shell gaps in spherical 
superheavy nuclei were discussed.

\begin{acknowledgements}
The work was supported by the
DoE grant DE-F05-96ER-40983.
\end{acknowledgements}

\end{document}